\documentclass[aps,twocolumn,showpacs,preprintnumbers,superscriptaddress]{revtex4-1}

\usepackage{array}
\newcolumntype{L}[1]{>{\raggedright\let\newline\\\arraybackslash\hspace{0pt}}m{#1}}
\newcolumntype{C}[1]{>{\centering\let\newline\\\arraybackslash\hspace{0pt}}m{#1}}
\newcolumntype{R}[1]{>{\raggedleft\let\newline\\\arraybackslash\hspace{0pt}}m{#1}}

\usepackage{amsmath}
\usepackage{amssymb}
\usepackage{amstext}
\usepackage{amsopn}
\usepackage{amsfonts}
\usepackage{amsxtra}
\usepackage[colorlinks]{hyperref}
\usepackage{url}
\usepackage{mathrsfs}
\usepackage[dvips]{graphicx}
\usepackage{relsize}
\usepackage{mathtools}
\usepackage{float}
\usepackage{bm}
\usepackage[dvipsnames]{color}
\numberwithin{equation}{section}

\newcommand{\mb}{\mathbf}

\newcommand{\h}{\hbar}

\newcommand{\kapa}{\boldsymbol{\kappa}}

\newcommand{\ee}{\varepsilon}
\newcommand{\kk}{\mathbf{k}}
\newcommand{\Q}{\mathbf{Q}}

\newcommand{\la}{\langle}
\newcommand{\ra}{\rangle}
\newcommand{\cd}{\cdot}
\newcommand{\ph}{\phantom\dag}

\newcommand{\e}{\varepsilon}
\newcommand{\s}{\sigma}

\newcommand{\om}{\omega}
\newcommand{\Om}{\Omega}
\newcommand{\al}{\alpha}

\newcommand{\de}{\delta}
\newcommand{\ka}{\kappa}
\newcommand{\D}{\Delta}

\newcommand{\E}{\mathscr{E}}

\def\mathclap#1{\text{\hbox to 0pt{\hss$\mathsurround=0pt#1$\hss}}}

\begin{document}
\def \brho{{\hbox{\boldmath $\rho$}}}
\def \beps{{\hbox{\boldmath $\epsilon$}}}
\def \bdelta{{\hbox{\boldmath $\delta$}}}

\title{ DC and optical signatures of the topological reconstruction of the Fermi surface  for electrons with parabolic band dispersion}
%
\author{Zoran Rukelj }
\email[]{zrukelj@phy.hr}
\author{Danko Radi\'{c} }
\affiliation{Department of Physics, Faculty of Science, University of Zagreb, Bijeni\v{c}ka 32, HR-10000 Zagreb, Croatia}
\date{\today}

\begin{abstract}

We study the main intra-band and inter-band transport properties at zero temperature
 of free electron-like system undergoing a topological reconstruction of the Fermi surface for the 
 two-dimensional and three-dimensional case. 
The calculated intra-band properties include the single-particle density of states, the total and the effective concentrations of electrons and the thermopower.
As for the inter-band case, the real part of the conductivity has been calculated within the vanishing inter-band relaxation approximation as a function of the incident photon energy.
 Within this approach, it is shown that the optical conductivity has a nonvanishing component parallel to the reconstruction wave vector 
 and the shape which depends on the value of the Fermi energy. 
 Each dimensionality has its particular features in the transport quantities presented in the paper, which are discussed and compared with those in the
 free electron scenario.
 Finally, we identify the signature of the  topological reconstruction of the Fermi surface in the intra-band and inter-band transport functions.
\end{abstract}

\maketitle

\section{Introduction}

The central goal of this paper is to identify signatures of the topological reconstruction of the  Fermi surface in the static  and dynamic electronic response functions in the free electron-like  two--dimensional (2D) and three-dimensional (3D) systems.
These response functions are easily experimentally accessible.
They are: the effective concentration of the intra-band charge carriers which defines the Drude weight, the thermoelectric power known as the Seebeck coefficient and 
the optical conductivity, namely its real part.
The emergence of the quantum phase transition associated with the topological reconstruction of the Fermi surface is predicted in the 2D and 3D nearly free electron gas systems \cite{Kad2018, Kad2019, Spaic202}. Here we mention a few notable examples of systems whose ground state is well enough approximated by the free electron dispersion. 

In the 2D case the recently proposed layered heterostructure LiBN \cite{Loncaric2018, Kim2018, Sumiyoshi2012, Rukelj2020a} has a single parabolic conduction band. The effective mass, the 
Fermi energy $E_F$ and correspondingly Fermi wave vector $k_F$ depend on the type of the alkali metal and on its concentration relative to the underlying BN hexagonal net. 
Another way of changing the Fermi energy is by the electrostatic doping \cite{S.2004} to which majority of the 2D materials are susceptible.
Furthermore, the real materials exhibiting the topological reconstruction of the Fermi surface, due to the spontaneous stabilization of the charge density wave ground state, are mostly effectively 2D systems such as the high-T$_c$ superconducting cuprates \cite{Keimer2015}, or certain intercalated graphite compounds 
\cite{Rahnejat2011}.

Contrary to 2D, a possible  3D system with parabolic electron dispersion for the reconstruction to take place is hard to find.  As it was shown \cite{Kad2018, Kad2019, Spaic202}, a necessary condition on the wave vector of reconstruction, relating the Fermi surfaces, is that it should be close to the integer multiple of $2k_F$. 
In the 3D metals the Fermi energy is changed by impurity doping. This in turn could invoke the nontrivial effects, leading to  the conduction band renormalisation near the Fermi energy.

In order to find the signature of the topological reconstruction of the Fermi surface in the electronic transport experiments, we
procede as follows:
First we define an auxiliary system which comprises of a free electron gas in the presence of a weak periodic crystal potential. 
This procedure is well know from the elementary solid state physics textbooks \cite{ashcroft} and in the vicinity of a single Bragg plane it  gives a two-band description of the electronic system.
For simplicity we adopt the simplest approximation where such a periodic potential has only one Fourier component. In the case of the uniaxial charge/spin density
wave this approximation is exact.
Hence, the resulting two-band ground state develops a pseudo-gap
thus defining boundary of the new Brillouin zone which now resembles 
 the infinitely long stripe (2D), or a cylinder (3D). 
Also, we can shift the Fermi energy relatively to the center of the pseudo-gap and calculate the effect of this shifting on the transport response functions which, 
to the knowledge of the authors, is not present in the literature.
Finally, the mechanism of topological reconstruction by the density wave is revised trough the minimization of the total electronic energy by formation of the self-consistent periodic potential.
A correspondence is then made between crystal and  self-consistent periodic potential, whose wave vector (which also determines the  Brillouin zone) is locked to the Fermi wave vector.
This makes our starting problem, topologically reconstructed electron gas, just a special case of the auxiliary model of free electrons in the weak perturbing potential. 

The paper is outlined as follows:

In Sec.~\ref{ham} we define the Hamiltonian along with the electron energies which are written in  dimensionless units defined on the  cylinder-like Brillouin zone.
In Sec.~\ref{dos} and \ref{dc} the single-particle density of state (DOS) is calculated for 2D and 3D case. Certain DOS features are to some extend visible in the 
effective concentration of electrons and in the Seebeck coefficient. Both quantities are calculated at near zero temperature and compared to the well-known free electron gas results. 
The real part of the optical conductivity is calculated using the Kubo formula in Sec.~\ref{optika}. 
We define and evaluate the inter-band current matrix element whose only non-vanishing  component is the one parallel to the wave vector of the density wave.
The closed form of the optical conductivity is found and its dependence on Fermi energy and dimensionality of the system analysed in details. 
Finally, the particular case of the phase with the topologically reconstructed Fermi surface is addressed.

\section{Two-band Hamiltonian}\label{ham}

The mathematical framework of this section does not depend on the system dimensionality. 
The generic mean-field  Hamiltonian  describes the free electron gas in the presence of the self-consistent, uniaxial reconstruction potential with amplitude $\D$
  and spatial modulation vector $\Q$
\begin{equation} \label{ham1}
  \hat{H} =    \sum_{\kk} \e_\kk c^{\dag}_\kk c^{\ph}_\kk  +  \D c^{\dag}_{\kk} c^{\ph}_{\kk-\mb{Q}} + \D c^{\dag}_{\kk-\mb{Q}} c^{\ph}_\kk.
\end{equation}
The electron dispersions have a parabolic shape $\e_\kk = c k^2$ where $c= \h^2/2m$, $\kk$ is electron wave vector, $m$ is the electron effective mass,  while the
second part in Eq.~(\ref{ham1}) is the coupling of electrons to the self-consistent reconstruction potential.
Writing the Hamiltonian Eq.~(\ref{ham1}) in its matrix form in the basis of $|\kk \ra$ and $| \kk-\Q \ra$ states
\begin{equation}\label{ham2}
 \mb{H} =  \begin{pmatrix}
 \e_{\kk} & \D  \\
 \D  & \e_{\kk-\mb{Q}}
\end{pmatrix}, 
\end{equation}
we notice the resemblance to the the Hamiltonian matrix found in the problem of determining the electron energies 
in the presence of a weak periodic crystal potential in the vicinity of a single Bragg's plane \cite{ashcroft}.
In that textbook example the crystal potential is assumed to have only a single Fourier component with the spatially-dependent form 
$2\D \cos( \mb{r} \! \cd \! \mb{Q})$. $\mb{Q}$ is the smallest reciprocal lattice vector.

   \begin{figure*}[tt!]
\includegraphics[width=.93\textwidth]{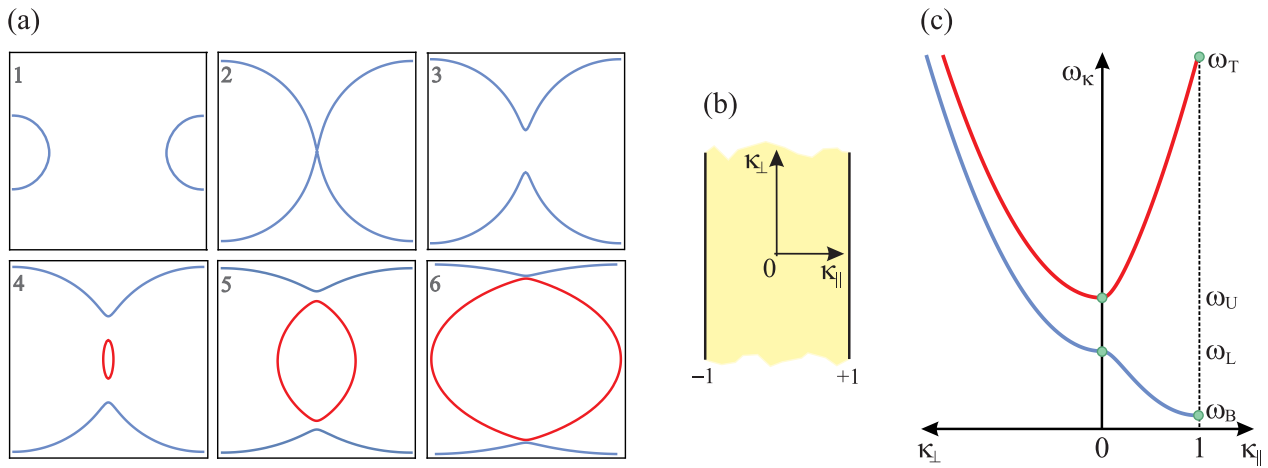} 
\caption{ (a) The Fermi surface for increasing values of the scaled Fermi energy $\om_F$. The blue and the red line correspond to the $s=-$ and $s=+$ band Eq.~(\ref{ham7}) respectively. In sub-figure 1 $\om_F$ is just above $\om_B$, in 2 $\om_F=\om_L$, in 3 $\om_F=\om_U$, 
in 4 $\om_F$ is just above $\om_U$, in 5 $\om_F$ is inbetween $\om_T$ and $\om_U$ and in 6 $\om_F$ is just below $\om_T$. (b) Schematic depiction of the Brillouin zone 
in dimensionless units of $\kappa$ after the reconstruction by a periodic potential. The zone is bounded in the $\kappa_\parallel$ direction and unbounded in the 
$\kappa_\perp$ direction. 
(c) The bands Eq.~(\ref{ham7}) along the direction of $\kapa$ are shown in (b) with four characteristic energy values marked by the green circles.}
\label{f1}
\end{figure*}

The diagonalisation of Eq.~(\ref{ham1}) is straightforward using the Bogoliubov unitary transformation 
 $\mb{U}\mb{H}\mb{U}^{-1} = \mb{E}$, where $\mb{E}$ is the diagonal eigenvalue matrix and $\mb{U}$ is \cite{Kupcic20041}
\begin{equation} \label{ham3}
\mb{U} =  \begin{pmatrix}
  \cos ({\vartheta_{\kk}}/{2}) & \sin ({\vartheta_{\kk}}/{2})
  \vspace{2mm} \\
  - \sin ({\vartheta_{\kk}}/{2})  &  \cos ({\vartheta_{\kk}}/{2})
 \end{pmatrix}.
\end{equation}
The auxiliary angle $\vartheta_\kk$  is a function of the Hamiltonian matrix elements in Eq.~(\ref{ham1})
\begin{equation} \label{ham4}
 \tan \vartheta_{\kk} = \frac{2\D}{\e_{\kk-\mb{Q}} - \e_{\kk}}. 
\end{equation}
Utilizing  Eq.~(\ref{ham3}) and   Eq.~(\ref{ham2}) in the above-described way, the Bloch energies are obtained. 
They are defined within the Brillouin zone with the periodicity determined by $\mb{Q}$
and are labeled by index $s \in \{ +,-\}$
\begin{equation}\label{ham5}
 E_\kk^{\pm} = \frac{1}{2} \left(\e_{\kk-\mb{Q}} +  \e_{\kk} \right) \pm \frac{1}{2} \sqrt{\left(\e_{\kk-\mb{Q}} - \e_{\kk} \right)^2 + 4\D^2 }.
\end{equation}
To make the mathematical treatment as simple as possible, three modifications  are done in electron dispersion Eq.~(\ref{ham5}).

First, $\kk$ is defined relatively to $\mb{Q}$. This way 
the Bloch wave vector may be decomposed as $\kk = \kk_\perp + \kk_\parallel$ with respect to the $\Q$ direction.

Further, the origin of the newly-formed Brillouin zone is shifted by $\kk \to \kk + \mb{Q}/2$.  That way the point of the band splitting shifts to the origin of Brillouin zone.
Implementing these two changes in Eq.~(\ref{ham5})  we get 
\begin{equation}\label{ham6}
 E_\kk^s = ck_\perp^2 +ck_\parallel^2+ c(Q/2)^2 + s \sqrt{ c^2k_\parallel^2Q^2 + \D^2 }.
\end{equation}
The final modifications defines the dimensionless variables, i.e. $\kapa = 2\kk/Q$,  and scaling the energies Eq.~(\ref{ham6}) to $\e_Q$,
\begin{equation}\label{ham7}
 \om_{\kapa}^s \equiv \frac{E_{\kapa}^s}{\e_Q} =   \kappa_\perp^2 +\kappa_\parallel^2+ 1 +s \sqrt{ 4 \kappa_\parallel^2 + \eta^2 },  
\end{equation}
 which are shown in Fig.~\ref{f1}. The energy scale $\e_Q = c(Q/2)^2$ in Eq.~(\ref{ham7})  is associated with the bare electronic energy at the Bragg's
 plane prior to the pseudo-gap opening. 
The dimensionless  parameter $\eta = \D/\e_Q$ is a measure of the strength of the perturbating potential. In the general perturbative crystal potential approach,
as well as in the case of the topological reconstruction, we expect $\eta \ll 1$.

The Brillioun zone, over which Eq.~(\ref{ham7}) is spanned, resembles an infinitely long
cylinder in $\kappa_{\perp} \in [-\infty, \infty]$ direction of total width $\kappa_{\parallel} \in [-1,1]$.
In the forthcoming calculation, there are four energy constants which are linked to the bands Eq.~(\ref{ham7}), of particular importance. 

The bottom ($B$) energy of the $s=-$ band and the top ($T$) energy of the $s=+$ band 
within the $\kappa_{\perp} = 0$ crossection of the Brillouin zone are located at the $\kappa_{\parallel} =1$ (see Fig.~\ref{f1})
\begin{equation}\label{ham8}
 \om_{\kappa_{\perp} = 0, \kappa_{\parallel} =1}^{\pm} \equiv \om_{T,B}  = 2 \pm \sqrt{ 4  + \eta^2 }. 
\end{equation}
Also, the values $\om^\pm_{\kapa}$ at the center of the Brillouin zone (the pseudo-gap region) $\kapa=0$ are important.
We label them by indices  $L$ and $U$ depending on their value
\begin{equation}\label{ham9}
 \om_{\kappa = 0}^{\pm} \equiv \om_{U,L}  = 1 \pm \eta,
\end{equation}
being the elliptic point in the upper and saddle point in the lower band, named the upper and the lower
"Lifshitz point" respectivelly \cite{ilja}.
So, the maximal vertical energy difference between the two bands is 
\begin{equation}\label{ham10}
 \om_T -\om_B = 2  \sqrt{ 4  + \eta^2 }, 
\end{equation}
while, correspondingly, the  width of the pseudo-gap is
\begin{equation}\label{ham11}
 \om_U -\om_L = 2 \eta.
\end{equation}
All the transport properties including the DOS to which we shall refer to in the next section, will be the piecewise functions of energy on the intervals defined by Eqs.~(\ref{ham8})--(\ref{ham11}).

\section{Density of states}\label{dos}

Here we calculate the single-particle DOS per unit volume for the 2D ($d=2$) and 3D ($d=3$) case. 
The mathematical procedure outlined in this section is used throughout the rest of the paper and is presented in detail in Appendix \ref{apa}.
DOS is defined as
\begin{equation}\label{dos1}
 G(E) = \frac{2}{V} \sum_{s, \kk}  \delta( E -E^s_\kk),
\end{equation}
with the  bands $E^s_\kk$ given by Eq.~(\ref{ham5}). 
 Changing the sum in Eq.~(\ref{dos1}) to an integral over  $\kapa$  and introducing the scaled energy $\om = E/\e_Q$, as defined in the previous section,  we get
\begin{eqnarray}\label{dos2}
  \hspace{0mm}G_d(\om)\, &&= \frac{2^{3-2d}}{\pi^2}  \frac{Q^d}{\ee_Q}  \sum_s \int_0^\infty  \ka^{d-2}_\perp   d\ka_\perp  \int_0^1  \,  d\ka_{\parallel}  \nonumber \\
 && \hspace{0mm} \times  \delta \left( \om - \kappa_\perp^2 - \kappa_\parallel^2 - 1 -s \sqrt{ 4 \kappa_\parallel^2 + \eta^2 } \, \right). 
\end{eqnarray}
The dimensionality $d$ enters explicitly in Eq.~(\ref{dos2}) in the prefactors, giving a correct unit of the $d$-dependent DOS,  and also as a parameter in the  integral
 over $\kappa_\perp$.  
Also, we have exploited the fact that Eq.~(\ref{ham7}) is an even function of $\kapa$.
The way to tackle the integral Eq.~(\ref{dos2}) which contains the Dirac delta function of another function $g(x)$ is to decompose the $\de$-function as a sum over the roots 
$x_0$
\begin{equation}\label{dos3}
 \de(g(x)) = \sum_{x_0} \frac{\de(x -x_0)}{\left| \partial g(x) / \partial x \big|_{x_0} \right| }, \hspace{3mm} g(x_0) = 0.
\end{equation} 
It is optimal to deal with the $\kappa_\perp$ variable first. Applying the formula  Eq.~(\ref{dos3}) to Eq.~(\ref{dos2}) we get
\begin{equation}\label{dos4}
G_d(\om)  =  \frac{2^{2-2d}}{\pi^2}  \frac{Q^d}{\ee_Q} \sum_s   \int_0^1   d\ka_{\parallel} \,  \{\kappa_\perp\}_0^{d-3}   \, \Theta \left( \{\kappa_\perp\}_0 \right), 
\end{equation}
where   $ \{\kappa_\perp \}_0$ is the real root of the argument of the delta function in Eq.~(\ref{dos2})
\begin{equation}\label{dos5}
 \{\kappa_\perp\}_0 =  \sqrt{\om - \kappa_\parallel^2- 1 -s \sqrt{ 4 \kappa_\parallel^2 + \eta^2 } }.
\end{equation}
The Heaviside step function $\Theta$ in Eq.~(\ref{dos4}) changes the integration limits (it is $1$ for positive arguments and zero otherwise). This is equivalent to the constraint that Eq.~(\ref{dos5}) is real and within the interval $ \{\kappa_\perp\}_0 \in [0, \infty]$.
This in turn imposes restrictions on the integration boundaries of $\ka_{\parallel}$ making them $(\om,s)$-dependent, transforming (\ref{dos4}) to 
\begin{equation}\label{dos6}
G_d(\om) =  \frac{2^{2-2d}}{\pi^2}  \frac{Q^d}{\ee_Q} \sum_s   \int_{b_s(\om)}^{t_s(\om)} \!\! \{\kappa_\perp\}_0^{d-3}  \, \, d\ka_{\parallel}    .
\end{equation}
In Appendix \ref{apa} it is shown how to obtain limits (the bottom $b_s(\om)$ and the top $t_s(\om)$) of integration.
Here we merely state their value:
\begin{eqnarray}\label{dos7}
 && {\rm{case}} \hspace{5mm} s=- \, : \nonumber \\
 && \om_B < \om < \om_L \Rightarrow  \begin{cases}
t_{-}(\om) = 1,\\
b_{-}(\om) =  \sqrt{1+\om - \sqrt{4\om + \eta^2}}
\end{cases} \nonumber \\
 &&  \om > \om_L \Rightarrow  \begin{cases}
t_{-}(\om) = 1,\\
b_{-}(\om) = 0
\end{cases} \nonumber \\
\vspace{3mm} \nonumber \\
 && {\rm{case}}  \hspace{5mm} s= + \, : \nonumber \\
 && \om_U < \om < \om_T \Rightarrow  \begin{cases}
t_{+}(\om) = \sqrt{1+\om - \sqrt{4\om + \eta^2}}\\
b_{+}(\om) =  0.
\end{cases} \nonumber \\
\end{eqnarray}
\begin{figure}[tt]
\includegraphics[width=.48\textwidth]{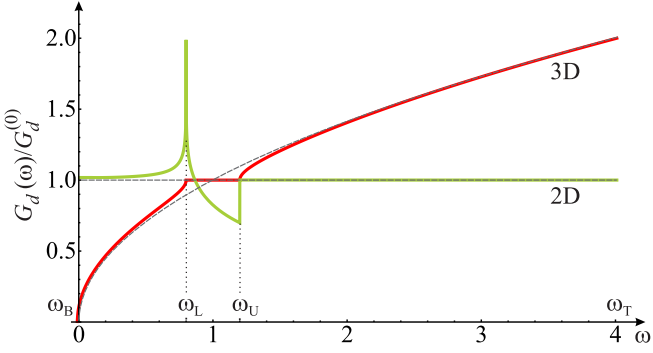} 
\caption{ The density of electron states (DOS) of a reconstructed system described by the bands Eq.~(\ref{ham7}) as a function of the scaled energy $\om$ in units 
of $G^{(0)}_d $ for $\eta=0.2$. Full green and red lines represent the 2D and 3D case respectively. The dashed lines represent the 2D and 3D free electron DOS Eq.~(\ref{dos8}). }
\label{f2}
\end{figure}

First we calculate the DOS for the free electron bands in the 3D and 2D case.
For consistency, the free electron DOS is also expressed in  terms of $\om$ and $\e_Q$
\begin{equation}\label{dos8}
G^{Free}_d(\om) = \frac{1}{2^{d+1}\pi^{d-1}} \frac{Q^d}{\e_Q} \om^{d/2-1} \equiv G^{(0)}_d \om^{d/2 -1},
\end{equation}
showing the usual constant or $\sqrt{\om}$-dependence in 2D and 3D,  respectively, as depicted in  Fig.~\ref{f2} by dashed lines.
The 3D  DOS Eq.~(\ref{dos6}) can be written down immediately since the integration is easily preformed  
giving 
\begin{eqnarray}\label{dos9}
&&\hspace{0mm} G_3(\om)/G_3^{(0)} =  \nonumber \\
&& \bigg( 1-  \sqrt{1+\om - \sqrt{4\om + \eta^2}}\bigg) \Theta(\om-\om_B)\Theta(\om_L - \om) +  \nonumber \\
&&\hspace{-0mm}  \, \Theta(\om - \om_L) +  \sqrt{1+\om - \sqrt{4\om + \eta^2}} \, \Theta(\om-\om_U)\Theta(\om_T - \om),  \nonumber \\
\end{eqnarray}
which is shown as red line in Fig.~\ref{f2}. Several features of this piecewise function stand out when compared to the free electron case. 
First, the onset of DOS is at
$\om_B$. 
Secondly, the emergence of the van Hove singularities at the points $\om_{L}$ and $\om_U$, as well as a constant value of the DOS between them,
$\om \in ( \om_L, \om_U )$, is in contrary to the overall $\sim \sqrt{\om}$ shape as anticipated by Eq.~(\ref{dos8}).
This constant value, as reported in \cite{Spaic202}, also deviating form the result stated in \cite{ashcroft}, has a profound influence on the DC transport properties 
as  shown in the next section.

For the 2D case we obtain the DOS numerically by inserting the boundaries Eq.~(\ref{dos7}) and the root Eq.~(\ref{dos5}) into Eq.~(\ref{dos6}).
The result is shown as a green line in Fig.~\ref{f2}. As in the previous 3D case here also we notice differences when compared with the constant DOS as predicted for the 
free electron model Eq.~(\ref{dos8}). 
The main distinction is visible for energies around the pseudo-gap region. At the energy $\om_L$ a logarithmic singularity is formed due to the saddle geometry of the band. 
 Once the $\om_U$ is reached, DOS jumps abruptly to  the free electron gas value Eq.~(\ref{dos8}) and continues so until $\om_T$ is reached.

\section{ Carrier concentrations and  low-$T$ thermopower}\label{dc}

For the upcomming analysis, we regard  the scaled Fermi energy $\om_F \equiv E_F/\e_Q$ as a variable. 
The way the Fermi energy is changed is not of our primary concern, nor shall we  go into the discussion about the possible influence that the doping procedure has on 
stability of the band structure. 
This provides an insight in the often-used charge transport quantities that
depend not only on the value of the Fermi energy, but also on the direction of the applied 
external perturbation.

First we calculate the total concentration of electrons $n_{tot}$ as it depends on the Fermi energy.
Second, the main component of the Drude weight, the effective concentration of conducting electrons $n_{\al}$, is calculated. The connection with the 
experiment here is simple since the Drude weight is measured in the reflectivity experiments (plasma edge). 
In a free electron-like system this concentration is trivially related with the total concentration of electrons \cite{Rukelj2020a}, but
 in the system described by more  "exotic" bands like Eq.~(\ref{ham6}), the two may differ significantly as seen in graphene for example \cite{Kupcic2014g}. 
Finally, the Seebeck coefficient or thermoelectric power $S_{\al}$ is calculated in the $T=0$ limit using the  Mott formula.

\begin{figure}[tt]
\includegraphics[width=.48\textwidth]{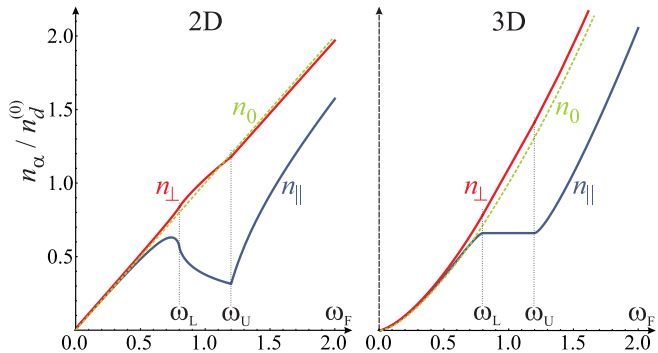} 
\caption{Total and effective concentration of electrons in 2D and 3D case as a functions of scaled Fermi energy $\om_F$ in units of $n^{(0)}_d$ Eq.~(\ref{n3}) 
for $\eta = 0.2$ and for $m=m_e$.
It is shown that total concentration is equal to the perpendicular effective concentration $n_{tot}=n_{\perp}$ in both dimensionalities. On the other hand $n_{\parallel}$ 
has features of its own.
The free electron concentration $n_0$ is represented by a green dashed line.}
\label{f3}
\end{figure}

\subsection{Total concentration of electrons $n_{tot}$}\label{ntot}

We start by writing  the total zero-temperature concentration of electrons in the momentum representation, 
\begin{equation}\label{n1}
 n_{tot}^{(d)}(E_F) =  \frac{2}{V} \sum_{s, \kk} \Theta(E_F - E^s_\kk ),
\end{equation}
where $V$ is the system volume, and in the scaled energy representation, 
\begin{equation}\label{n2}
 n_{tot}^{(d)}(\om_F) = \e_Q \int_0^{\om_F} G_d(\om) d\om,
\end{equation}
where, in the later,  the recently calculated DOS has been utilized.
The $ n_{tot}^{(d)}$, Eq.~(\ref{n2}) is a function  of scaled Fermi energy $\om_F$
and it is depicted in the Fig.~\ref{f3} for 2D and 3D case ($n_{tot}^{(d)}=n_{\perp}^{(d)}$ because of the reasons given in the following subsection). 
On the same figure a $d$-dependent free electron value of electron concentration 
\begin{equation}\label{n3}
 n_0(\om_F) =  \frac{Q^d}{ 2^{d}\pi^{d-1} d}  \om_F^{{d}/{2}} \equiv n^{(0)}_d \om_F^{{d}/{2}}
\end{equation}
is drawn by the green dashed lines. 

The common feature in both 2D and 3D is that the total concentration Eq.~(\ref{n2}) and its free electron analog Eq.~(\ref{n3}) are almost equal for energies
in the interval $(\om_B, \om_L)$. Above $\om_L$ they start to deviate, slightly more in 3D then in 2D.

\subsection{Effective concentration of electrons $n_{\al}$}\label{neff}

Here, the effective concentrations of electrons $n_{\al}$ that participate in the DC transport are calculated. Index $\alpha$ denotes a Cartesian component
with respect to the reconstruction wave vector, i.e. $\al  \in (\parallel, \perp)$,
of the effective concentration. The DC conductivity is defined as $\s^{DC} = e^2 \tau n_{\al}/m_e$ where $\tau$ is the scattering relaxation time, $m_e$ is 
a bare electron 
mass and the zero-temperature $(T=0)$ expression for  $n_{\al}$  \cite{Kupcic20144} is given in two equivalent forms
%
%
\begin{subequations}
    \label{nef1}
    \begin{align}
         n^{(d)}_\al(E_F) &= \frac{2}{V}\frac{m_e}{\h^2}  \sum_{s,\kk}  \left( \frac{\partial E^s_\kk}{\partial k_\alpha}\right)^2 \delta( E_F -E^s_\kk) \mbox{,}
                \label{a-prva} \\
        &= \frac{2}{V} \frac{m_e}{\h^2}  \sum_{s,\kk}  \frac{\partial^2 E^s_\kk}{\partial k_\alpha^2} \, \Theta( E_F -E^s_\kk) \mbox{.}
                \label{b-druga}
    \end{align}
\end{subequations}
The two expressions in Eq.~(\ref{nef1}) are obtainable from one another by partial integration. Here, as well,  $n_{\al}$ will  be expressed in
terms of the dimensionless
units $\om_F$.

The two distinct directions $\al  \in (\parallel, \perp)$ in electron dispersions Eq.~(\ref{ham6}) imply the difference between $n_\perp$ and $n_\parallel$.
It is easy to check that $\partial^2 E^s_\kk /\partial k_\perp^2 = \h^2/m$ and thus Eq.~(\ref{b-druga}) is the same as Eq.~(\ref{n1}), $n_{tot}^{(d)} = n_\perp^{(d)}$,
if $m=m_e$, which holds for the $\al = \perp$ case regardless 
of dimensionality $d$. 

On the other hand, this is not so for the $\al = \parallel$ case. The second derivative of the electron dispersion Eq.~(\ref{ham6}) over $k_{\parallel}$ is not a constant, but a rather complicated function of $k_{\parallel}$. Since we have already presented  the solution for the  integrals involving the $\delta$-function,
we shall proceed by evaluating $n_{\parallel}$ using Eq.~(\ref{a-prva}). 
 Changing the sum  into an integral over $\kapa$ and introducing $\om$ as defined in the previous section, the following result is obtained
\begin{eqnarray}\label{nef2}
 && \hspace{0mm}n^{(d)}_{||}(\om) =  \frac{2^{4-2d}}{\pi^2} Q^d \sum_s \int_0^\infty  \ka^{d-2}_\perp   d\ka_\perp  \int_0^1 \ka^2_{\parallel} \,  d\ka_{\parallel}  \nonumber \\
 && \hspace{0mm}  \bigg(1+ \frac{2s}{\sqrt{ 4 \kappa_\parallel^2 + \eta^2 }}  \bigg)^2 \delta \left(\kappa_\perp^2 +\kappa_\parallel^2+ 1 +s \sqrt{ 4 \kappa_\parallel^2 + \eta^2 } - \om \right), \nonumber \\
\end{eqnarray}
which, after the $\delta$-function evaluation by the same recipe from the Sec.~\ref{dos}, 
gives 
\begin{eqnarray}\label{nef3}
 && \hspace{0mm}n^{(d)}_{||}(\om) = \frac{2^{3-2d}}{\pi^2} Q^d \, \sum_s     \nonumber \\
 && \hspace{0mm} \int_{b_s(\om)}^{t_s(\om)} \ka^2_{\parallel} \,  d\ka_{\parallel}  \bigg(1+ \frac{2s}{\sqrt{ 4 \kappa_\parallel^2 + \eta^2 }}  \bigg)^2  \{\kappa_\perp\}_0^{d-3}.
\end{eqnarray}
The differences between the effective  $n^{(d)}_{||}$ and the total $n^{(d)}_{tot}$, concentration are shown in the Fig.~\ref{f3}. Unlike the $n^{(d)}_{tot}$ 
or equivalently $n^{(d)}_{\perp}$, $n_{||}^{(d)}$ is extremely susceptible to the 
features originating from the pseudo-gap.
Several features are highlighted depending on the position of $\om_F$.
The main feature of 2D effective concentration  is the "shark fin" shape at $\om_F$ within the pseudo-gap.
The logarithmic divergency in DOS at energy of the lower Lifshitz point,  $\om_L$ (the saddle point in the lower band), corresponds to the inflection point in $n^{(2)}_{||}(\om_L)$. 
Also, the discontinuity in the DOS 
generates the discontinuity in the slopes of $n^{(2)}_{||}(\om_U)$ at the energy of the higher Lifshitz point, $\om_U$ (bottom of the upper band). 
Above the pseudo-gap,   $n^{(2)}_{||} \sim \om_F$ 
gradually tends to Eq.~(\ref{n3}) as $\om_F$ increases.

The features of the effective concentration in 3D system are more "tamed" than those in 2D. Outside the pseud-gap,  $n^{(3)}_{||}(\om_F)$ has roughly
a $\om_F^{3/2}$-dependence as shown by Eq.~(\ref{n3}).  Within the pseudo-gap region, $n^{(3)}_{||}(\om_F)$ is a constant.

It is particularly visible on the example of this system how $n_{tot}^{(d)}$ and $n_{\al}^{(d)}$ differ. Although  $n_{\perp}^{(d)} = n_{tot}^{(d)}$, they are both aproximately by 
a factor of two larger then $n_{\parallel}^{(d)}$ for $\om_F > \om_L$. Eventually they meet at higher values of $\om_F$. The discrepancy is a result of an unphysically large value of $\eta$, chosen as such merely for the matter of presentation, for depiction of the two concentrations. This discrepancy also serves as a reminder to the fact that even a simple distortion like the pseudo-gap opening on the Fermi surface
changes the concentration of electrons participating in the DC conductivity substantially comparing to their total number. As $\eta$ is set to zero,
$n_\al$ and $n_{tot}$ naurally become equal to $n_0$.

\subsection{The Seebeck coefficient $S_{\al}$}\label{seebak}

\begin{figure}[tt]
\includegraphics[width=.48\textwidth]{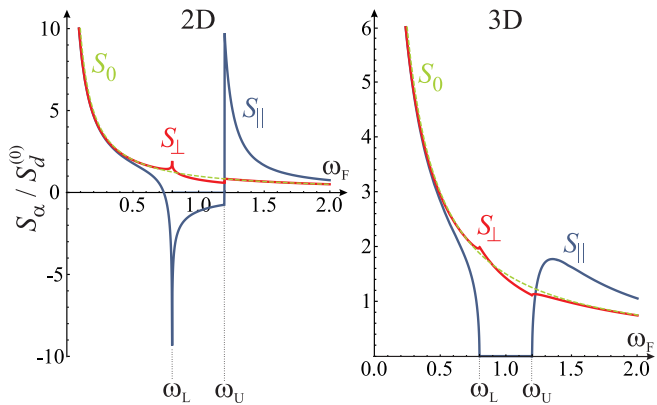} 
\caption{ Direction-dependent Seebeck coefficient, $S_{\parallel}$ and $S_{\perp}$, as a function of scaled Fermi energy $\om_F$ for the 2D and 3D system described by Eq.~(\ref{ham7})
in units of $S_d^{(0)}$ for $\eta=0.2$ and $m=m_e$. The free electron gas value $S_0$ Eq.~(\ref{seb2}) is presented by the green dashed line. }
\label{f4}
\end{figure}

We use the  well-known Mott formula for  $S_{\al}$ \cite{soljom, ashcroft}. 
 This formula is a result of the Sommerfeld  expansion ($k_BT \ll E_F$) of the Onsanger's transport coefficients and it reads
\begin{equation}\label{seb1}
 S_{\al}(\om_F) \approx \frac{\pi^2 k_B^2T}{3e \e_Q n_\al(\om_F)} \frac{\partial n_\al(\om) }{\partial \om}  \bigg|_{\om_F}.
\end{equation}
As noted in the in Sec.~\ref{neff}, $n_{\al}$ depends differently in $\al = \parallel$ and $\al = \perp$ case, so consequently  will 
$S_{\al}$.
For comparison, the Seebeck coefficient for the 2D and 3D free electron gas is derived using Eqs.~(\ref{seb1}) and (\ref{n3}).
This gives simply 
\begin{equation}\label{seb2}
 S_0(\om_F) = \frac{\pi^2 k_B^2T}{3e\e_Q} \frac{d}{2\om_F} =  \frac{S_d^{(0)}}{\om_F},
\end{equation}
since, as noted before for the free electron gas, the total and the effective concentrations are equal. Expression (\ref{seb2}) is shown in Fig.~\ref{f4}
as a green dashed line.
The results for the Seebeck coefficient Eq.~(\ref{seb1}) with two different components $\al \in (\perp, \parallel)$ are shown in Fig.~\ref{f4} for the 2D and 3D case.
The common characteristic for 2D and 3D case is that, for Fermi energies small compared to $\om_L$, $S_{\al}^{(d)}$ is equal to the free electron value Eq.~(\ref{seb2}).
Also, in general, $S_{\perp}^{(d)}$ deviates weakly from the free electron result with the main differences located around and in the pseudo-gap region.
$S_{\parallel}^{(d)}$ on the other hand has a rich structure. For the 2D system, it has a logaritmic divergency at $\om_L$ and a discontinuity at $\om_U$. Also, between these two points $S^{(2)}_{\parallel}$ changes sign.
In the 3D case $S_{\perp}^{(3)}$ manages to follow free electron result with minor deviations in the form of small spikes at $\om_L$ and $\om_U$. 
On the other hand, $S_{\parallel}^{(3)}$ vanishes in the pseudo-gap. This is due to the constant DOS Eqs.~(\ref{dos9}) for this energy
interval.
In the $\om_F \gg \om_L$ limit, as well as for $\eta \to 0$, $S_\al(\om_F) \to S_0(\om_F)$, $\forall \alpha$, both in the 2D and 3D case.

\section{Optical conductivity}\label{optika}

Here we calculate the zero-temperature optical response of the electron system described by Eq.~(\ref{ham1}) in 2D and 3D case. 
The calculation consists of defining the optical conductivity tensor within the two-band picture in the limit of vanishing inter-band relaxation and with the 
current matrix elements derived from the Hamiltonian Eq.~(\ref{ham1}). We shall see that shape of the real part of the optical conductivity strongly depends on the 
values of the scaled Fermi energy $\om_F$, and weakly on the dimensionality $d$ of the system.

\subsection{Optical conductivity Kubo formula}

In the two-band model the complex inter-band conductivity tensor is defined  \cite{Rukelj20211} as a function of the incident photon energy $\E$
\begin{equation}\label{op1}
 \s_{\al}(\E) = \frac{2i\hbar}{V}  \sum_{s \neq s'= \pm} \sum_{\kk} \frac{|J^{ss'}_{\al \kk}|^2}{E^s_{\kk}-E^{s'}_{\kk}} \frac{f^{s'}_\kk - f^s_\kk}{\E-E^s_{\kk}+E^{s'}_{\kk} +i\Gamma},
\end{equation}
where $\al$ is index of the Cartesian component of the inter-band conductivity tensor. The only $\al$-dependent part in the conductivity formula Eq.~(\ref{op1}) is the inter-band current matrix element $J^{ss'}_{\al \kk}$.  These elements are  part of the diagonalized current matrix defined as a unitary transformation of the  Hamiltonian
Eq.~(\ref{ham2}) matrix derivative 
\begin{equation} \label{op2}
  \mb{J}_{\al} =  \frac{e}{\hbar}\mb{U} ( \partial \mb{H}/  \partial k_\al ) \mb{U}^{-1},
\end{equation}
with the unitary matrix $\mb{U}$ given by Eq.~(\ref{ham3}).
In the expression Eq.~(\ref{op1}) $\Gamma$ is a small phenomenological relaxation parameter and 
in the limit $\Gamma \to 0$ the real part of the conductivity tensor for the incident photon energies $\E >0$ reduces to 
\begin{eqnarray}\label{op3}
&& {\rm{Re}} \,  \s_{\al}(\E) =  \frac{2 \hbar\pi}{\E V} \sum_{\kk} |J^{+-}_{\al,\kk}|^2 \big(f^-_\kk - f^+_\kk \big)\delta(\E -E^+_\kk + E^-_\kk ). \nonumber  \\
\end{eqnarray}
Once the elements Eq.~(\ref{op2}) are derived we can use Eq.~(\ref{op3}) for further analytical derivation.

\subsection{Optical conductivity of 2D and 3D system}\label{optika-kar}

We begin by evaluating the inter-band current element. Inserting the unitary matrix elements Eq.~(\ref{ham3})  and the derivatives of the Hamiltonian (\ref{ham2}) 
in Eq.~(\ref{op2})  we get  
\begin{eqnarray}\label{op4}
J_{\al \kk}^{+-} =  \frac{e}{2\h}\sin \vartheta_{\kk}  \frac{\partial (\e_{\kk-\Q} - \e_{\kk})}{\partial k_{\alpha}}.
\end{eqnarray}
The only non-vanishing component in the derivative of the free electron dispersion (Sec.~\ref{ham})  in the above expression is $\al = \parallel$. Therefore, the optical conductivity in the presented model has only the $\al = \parallel$ component.
Written in terms of  dimensionless variables $\kapa$ and $\eta$  the inter-band current element is 
\begin{eqnarray}\label{op5}
J_{\parallel \kapa}^{+-} =  \frac{e}{\h} Qc  \frac{\eta}{\sqrt{4\kappa^2_{\parallel}+ \eta^2 }}.
\end{eqnarray}
We shall omit the label $\parallel$ when  addressing the real part of the optical conductivity which we calculate by
inserting the Eq.~(\ref{op5})  into Eq.~(\ref{op3})  and changing the sum to an integral over the dimensionless variable $\kapa$ within the limits given in Sec.~\ref{ham}.
As a general result for dimensionality $d$ we find
\begin{eqnarray}\label{op6}
&& {\rm{Re}} \,\s^{(d)}(\Om) = \frac{2^{9-2d}}{\pi} \s_0Q^{d-2} \frac{\eta^2}{\Om} \int_0^\infty \kappa_\perp^{d-2}d \kappa_\perp \int_0^1  d\kappa_{\parallel} \times \nonumber \\
 &&\frac{\Theta \left(E^-_\kappa - E_F\right)-\Theta \left(E^+_\kappa - E_F \right)}{4\kappa^2_{ \parallel}+ \eta^2  }  \de \left( \Om  - 2\sqrt{4\kappa^2_{\parallel}+ \eta^2 } \right). \nonumber \\
\end{eqnarray}
In writing the above integral we have used the $T=0$  Fermi-Dirac distribution function  $f(E_\kappa) =\Theta(E_F- E_\kappa )$. 
Also, in the above expression,  the scaled dimensionless variable for the incident photon energy, $\Om = \E/\e_Q$, is 
introduced alongside the conductivity constant $\s_0 = e^2/4\h$. 
 
\begin{figure*}[tt]
\includegraphics[width=.98\textwidth]{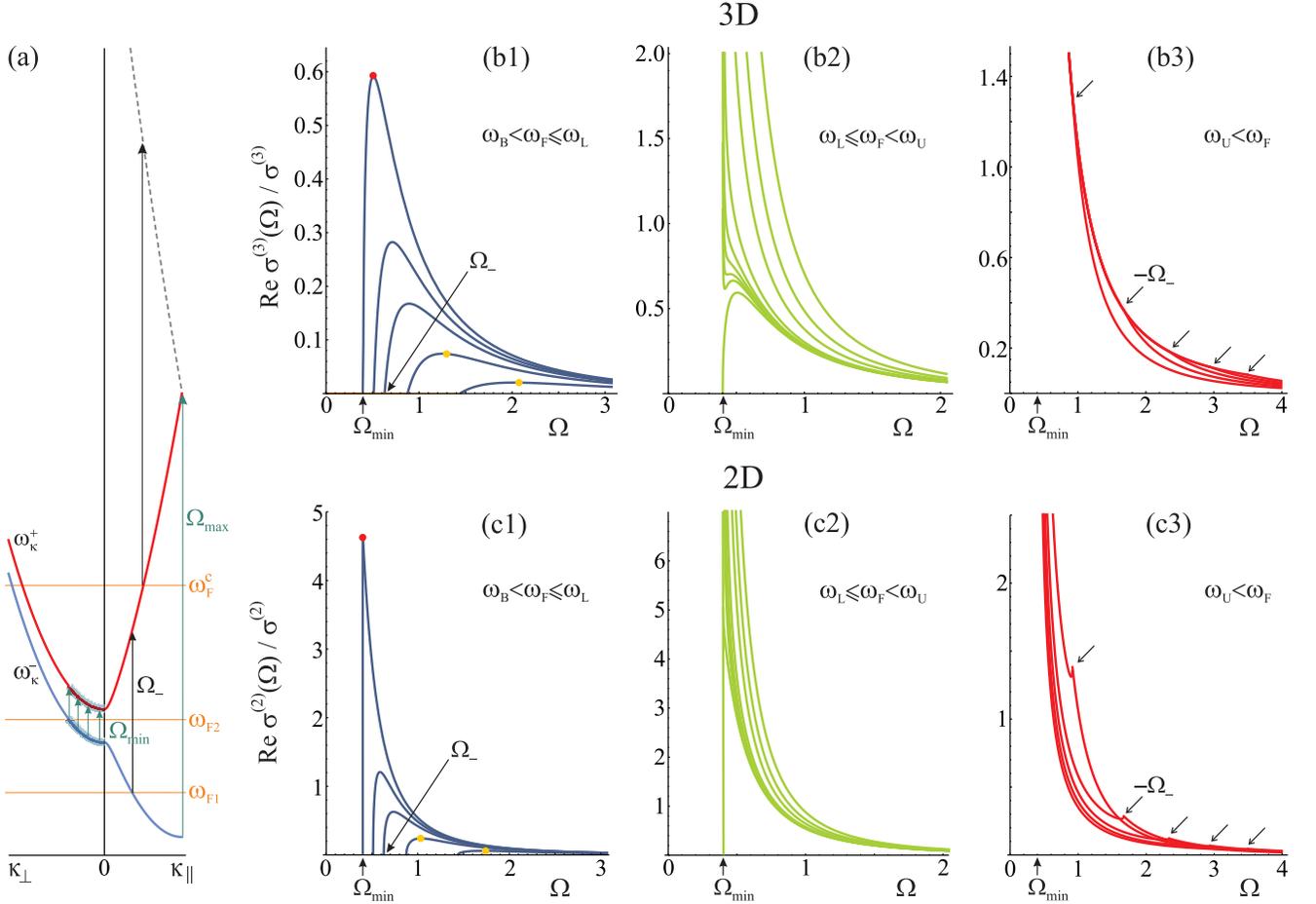} 
\caption{ (a) Schematic view of the characteristic regions for optical excitations with respect to the band structure. The position of several characteristic Fermi energies $\om_F$ is shown in orange 
alongside with $\Om_{min}$, $\Om_{max}$ and a optical threshold energy $\Om_-$ with respect to $\om_{F1}$. The thick line segments on the bands around $\om_{F_2}$ indicate filled and empty 
states that are shifted with respect to one another by the energy $\Om_{min}$. A hypothetical third band is shown as a dotted line together with the critical $\om^c_F$.
Gray thick lines and arrows on the bands around $\om_{F2}$ indicate states excited from the $\om_{\kappa}^-$ to $\om_{\kappa}^+$ band by a constant energy $\Om_{min}$
leading to the divergent optical response.
(b1)--(c3) ${\rm{Re}} \, \s^{(d)}(\Om)$ plotted for several values of $\om_F$ within the three intervals as described in the main text. The conductivity is given in units 
of $\s^{(d)}$ with parameter $\eta=0.2$ defining the position of $\Om_{min}$ as indicated by an arrow. The characteristic points such as the onset of optical excitation at $\Om_-$,  the positions of maxima (at energy $\Om_0$) and the kinks position at $-\Om_-$ are indicated on the figures by dots or arrows. For comparison, both 3D and 2D ${\rm{Re}} \, \s^{(d)}(\Om) $ are plotted
for the same values of $\om_F$ belonging to the one of three intervals as noted in figures.
In (b1) and (c1) ${\rm{Re}} \, \s^{(d)}(\Om) $ is plotted for  $\om_F = 0.4, \, 0.6, \,0.7,\, 0.75, \, 0.8 $. In (b2) and (c2) ${\rm{Re}} \, \s^{(d)}(\Om) $ is 
plotted for $\om_F = 0.8001, \, 0.805, \,0.807,\, 0.81, \, 0.815,\, 0.83,\, 0.9,\, 1,\, 1.2 $. In (b3) and (c3) ${\rm{Re}} \, \s^{(d)}(\Om) $ is plotted for 
$\om_F = 1.5, \, 2, \,2.5,\, 3, \, 3.5 $.}
\label{f5}
\end{figure*}

 In  Eq.~(\ref{op6}) the variable $\kappa_{ \parallel}$  is used for decomposing the $\delta$-function over its roots. According to the procedure from Sec.~\ref{dos} 
\begin{equation}\label{op7}
 \de \left( \Om  - 2\sqrt{4\kappa^2_{\parallel}+ \eta^2 } \right) = \frac{\Om}{8\{\kappa_{ \parallel}\}_0} \de\left(\kappa_{ \parallel}- \{\kappa_{ \parallel}\}_0  \right),
\end{equation}
where $\{ \kappa_{ \parallel}\}_0 = \sqrt{\Om^2 - (2\eta)^2}/4 $. 
The initial restriction  $ \kappa_{ \parallel} \in [0,1]$ 
limits the range of $\Om$ to 
\begin{equation}\label{op8}
  2\eta= \Om_{min} < \Om < \Om_{max}= 2\sqrt{4+\eta^2},
\end{equation}
and so does the interval on which ${\rm{Re}} \,\s^{(d)}(\Om)$ is defined.  The effect, that the restriction Eq.~(\ref{op8})
has on ${\rm{Re}} \,\s^{(d)}(\Om)$, can be summarized to ${\rm{Re}} \,\s^{(d)}(\Om)\propto \Theta(\Om-\Om_{min})\Theta(\Om_{max} -\Om)$.
Limits in Eq.~(\ref{op8}) are thus easily identified. 
The lower value, $\Om_{min}$, is the minimum vertical spacing between the valence and conduction band and is equal to the pseudo-gap width (see Figs.~\ref{f1}(c), \ref{f5}(a) and Eq.~(\ref{ham8}). 
The larger value, $\Om_{max}$,  is the maximum distance from the bottom of the valence to the top of the conduction band (see Figs.~\ref{f1}(c), \ref{f5}(a) and Eq.~(\ref{ham9})).

Inserting the $\de$-function Eq.~(\ref{op7}) back into  Eq.~(\ref{op6}) and changing $\kappa_{ \parallel} \to \{ \kappa_{ \parallel}\}_0 $ in the electron energies
$E^{\pm}_{\kapa}$ in arguments of the $\Theta$-functions, we have

\begin{eqnarray}\label{op9}
 && \hspace{-5mm}{\rm{Re}} \, \s^{(d)}(\Om) =  \nonumber \\
 && \frac{2^{10-2d}}{\pi} \s_0Q^{d-2} \eta^2  \frac{\Theta(\Om-\Om_{min})\Theta(\Om_{max} -\Om)}{\Om^2\sqrt{\Om^2-(2\eta)^2}} \times \nonumber \\
 &&  \int_0^\infty \kappa_\perp^{d-2}d \kappa_\perp  \bigg\{ \Theta \left[  \om_F - \frac{(\Om -4)^2 - (2\eta)^2}{16}  -\kappa^2_\perp    \right] \nonumber \\
 && \hspace{10mm}- \, \Theta \left[  \om_F - \frac{(\Om +4)^2 - (2\eta)^2}{16}-\kappa^2_\perp      \right] \bigg\}.
\end{eqnarray}
The $\Theta$-functions within the integral in Eq.~(\ref{op9}) are finite (equal to 1) only if their arguments are positive.
We exploit that by introducing the auxiliary function 
\begin{equation}\label{op10}
 R_\pm  = \sqrt{\om_F - \frac{(\Om \pm 4)^2 - (2\eta)^2}{16}  },
\end{equation}
so we can factorise the arguments of the  $\Theta$-functions in Eq.~(\ref{op9}) as
\begin{eqnarray}\label{op101}
 && \Theta \big(  R^2_\pm  - \kappa_\perp^2 \big) = \Theta \big(  (R_\pm  - \kappa_\perp)(R_\pm  + \kappa_\perp) \big) \nonumber \\
 &&  \equiv \Theta \big(  R_\pm  - \kappa_\perp \big).
\end{eqnarray}
In Eq.~(\ref{op101}) it states: if $R^2_\pm <0$ ($R_\pm $ is not real), then the argument of the $\Theta$-function on the left-hand side of Eq.~(\ref{op101}) 
is zero since $\kappa_{\perp} >0$  and the $\Theta$-function vanishes, as does the entire Eq.~(\ref{op9}). 
If $R^2_\pm  > 0$  ($R_\pm $ is real), the argument of the $\Theta$-function on the left-hand side of Eq.~(\ref{op101}) is positive and can be written as a product of 
monomials.
Since  $R_\pm  + \kappa_\perp >0$,  it does not influence the value of the $\Theta$-function, leaving effectively the result in Eq.~(\ref{op101}).
Thus, from Eq.~(\ref{op101}) we see that the $\Theta$-function influences only the upper limit of integration
\begin{eqnarray}\label{op11}
 && {\rm{Re}} \, \s^{(d)}(\Om) = \nonumber \\
 && \frac{2^{10-2d}}{\pi} \s_0Q^{d-2} \eta^2  \frac{\Theta(\Om-\Om_{min})\Theta(\Om_{max} -\Om)}{\Om^2\sqrt{\Om^2-(2\eta)^2}} \times \nonumber \\
 && \hspace{5mm} \bigg(  \int_0^{R_- } \kappa_\perp^{d-2}d \kappa_\perp - \int_0^{R_+ } \kappa_\perp^{d-2}d \kappa_\perp  \bigg).
\end{eqnarray}

Whether  Eq.~(\ref{op10}) is real or not and consequently  Eq.~(\ref{op11}) finite or zero,  
depends on the value of $\Om$ and $\om_F$. Of course, the allowed values of $\Om$ should fall within
the limits set by  Eq.~(\ref{op8}).
With a modest effort in determining the sign of the sub-root functions in  Eq.~(\ref{op10}), the following conclusions can be made for each integral in 
Eq.~(\ref{op11}) separately.

 $R_-$ is real (the sub-root function Eq.~(\ref{op10}) is positive) and hence the first integral in the curly brackets in  Eq.~(\ref{op11}) is finite
 for $\Om$ within the interval
 \begin{equation}\label{op12}
  \Om_{-}  < \Om <  \Om_+,
\end{equation}
where 
 \begin{equation}\label{op13}
   \Om_\pm = 4  \pm \sqrt{16\om_F + (2\eta)^2}.  
\end{equation}
Also, comparing  Eq.~(\ref{op13}) and  Eq.~(\ref{op8}) we conclude 
\begin{eqnarray}\label{op14}
 && \om_F < 1-\eta \,\, \,   \Rightarrow \,\,   \Om_- > \Om_{min}, \nonumber \\
  && \om_F > 1-\eta \,\, \,   \Rightarrow \, \,  \Om_- < \Om_{min},
\end{eqnarray}
and  $ \Om_+ > \Om_{max}$.

On the other hand, $R_+$ is real (the sub-root function Eq.~(\ref{op10}) is positive) and hence the second integral in the curly brackets in  Eq.~(\ref{op11}) is finite 
for $\Om$ in the interval
 \begin{equation}\label{op15}
  -\Om_{+}  < \Om <  -\Om_-.
\end{equation}
From Eq.~(\ref{op13}) we see that regardless of $\om_F$, $ -\Om_+ < \Om_{min}$, while $ -\Om_- > \Om_{min}$ if $\om_F> 1+\eta$. 

When collected, the conditions and restrictions (\ref{op12})---(\ref{op15}) are best expressed by the aid of the step function $\Theta$.
Defining a $d$--dependent constant of conductivity as
\begin{equation}\label{op151}
\s^{(d)} = \frac{2^{10-2d}}{\pi} \s_0 \eta^2  \frac{Q^{d-2}}{d-1},
\end{equation}
 we arrive to the final expression for the real part of the  $d$--dimensional optical conductivity
\begin{eqnarray}\label{op16}
 && {\rm{Re}} \, \s^{(d)}(\Om) = \s^{(d)}\frac{\Theta(\Om_{min}-\Om)\Theta(\Om-\Om_{max})}{\Om^2\sqrt{\Om^2-\Om^2_{min}}}\times   \nonumber \\
 && \Big[ R_-^{d-1} \Theta(\om_F-\om_L)\Theta( \Om_- - \Om)+ R_-^{d-1}\Theta(\om_L-\om_F)   \nonumber \\
 && \hspace{3mm} +  R_+^{d-1} \Theta(\Om + \Om_-)  \Theta(\om_U-\om_F) \Big].
\end{eqnarray}

There are two main features in Eq.~(\ref{op16}). The first one is the prefactor  which, regardless of $d$,  has a strong $\Om^{-3}$-dependency.
The second one is the part within the square brackets. Besides on $\Om$, it is also depends on $d$ and $\om_F$.
To fully understand how ${\rm{Re}} \, \s^{(d)}(\Om)$ evolves with $\om_F$ we analyse each dimensionality separately.

\subsection{3D optical conductivity features}

Here we describe the 3D $\Om$-dependent conductivity ${\rm{Re}} \, \s^{(d)}(\Om)$ as it depends on $\om_F$.\
\\

 Case $\om_F \leq \om_L$: \\
In  Fig.~\ref{f5}(b1) ${\rm{Re}} \, \s^{(3)}(\Om)$ is shown for several values of $\om_F$ bounded by condition $\om_F \leq \om_L$. 
The onset of the inter-band excitation begins at $\Om_-$ Eq.~(\ref{op3}), indicated by an arow in Fig.~\ref{f5}(b1) for a specific value of $\om_F = 0.7$, 
 and  ends at $\Om_{max}$.
 $\Om_-$ is the vertical distance between the bands as shown schematically in Fig.~\ref{f5}(a). In this range of $\om_F$ values, ${\rm{Re}} \, \s^{(3)}(\Om) $ is a humped curve expanding towards $\Om_{min}$, with increasing amplitude and spiking as $\om_F \to \om_L$.
For small enough $\om_F$, the position of maximum is roughly estimated to be at
 \begin{equation}\label{op17}
  \Om_0  \approx 8-4\sqrt{1+3\om_F},
\end{equation}
represented by orange dots in Fig.~\ref{f5}(b1).
Eq.~(\ref{op17}) was obtained under the assumption $\Om_- \gg \Om_{min}$. Clearly this breaks as we approach $\om_F \to \om_L$.

There is a steep linear dependence ${\rm{Re}} \, \s^{(3)}(\Om) \propto (\Om- \Om_-)/\sqrt{\Om_- - \Om_{min}}$  for photon energies just above the onset of inter-band excitations at $\Om_-$ (Fig.~\ref{f5}(b1)). The steepness increases as $\om_F \to \om_L$. 
Finally, for $\om_F = \om_L$  Eq.~(\ref{op16}) reduces to 
 \begin{equation}\label{op18}
  {\rm{Re}} \, \s^{(3)}(\Om) \approx  \frac{\s^{(3)}}{2^{3/2} \Om_{min}^{5/2}} \sqrt{\Om- \Om_{min}} \, (\Om_+ - \Om),
\end{equation}
for energies $\Om$ just above $\Om_- = \Om_{min}$. At the particular value, $\om_F = \om_L$, we can analytically find the position of the maximum which is located at
$\Om_0 = \Om_{min}(1+\sqrt{17})/4$, clearly above the threshold energy $\Om_{min}$ with the peak height of 
${\rm{Re}} \, \s^{(3)}(\Om_0) \approx 0.0134 \s_0^{(3)}(8-2.281 \Om_{min})/\Om_{min}^2$, depicted by the red circle in Fig.~\ref{f5}(b1).
The square-root in ${\rm{Re}} \, \s^{(3)}(\Om) \propto  \sqrt{\Om- \Om_{min}}$ for $\om_F =\om_L$ has its origin in the shape of DOS below $\om_L$ and above $\om_U$ which has a square-root dependence (see Eq.~(\ref{dos9}) and Fig.~(\ref{f2})).\
\\

 Case $\om_L \leq \om_F \leq \om_U$: \\
For  $\om_F$ slightly above $\om_L$, two features take place in ${\rm{Re}} \, \s^{(3)}(\Om)$. The first one is 
a one over a square-root divergency ${\rm{Re}} \, \s^{(3)}(\Om) \propto 1/\sqrt{\Om- \Om_{min}}$ for $\Om \gtrapprox \Om_{min}$, which quickly falls off 
to the  second feature, which is a residuum of the hump shown in Fig.~\ref{f5}(b2). 
As $\om_F$ increases and enters into the pseudo-gap, the hump in ${\rm{Re}} \, \s^{(3)}(\Om)$ shifts to the left  and eventually disappears leaving only a one over  square-root singularity
which is present for all $\om_F > \om_L$. The origin of the divergency can be easily traced from the Fig.~\ref{f5}(a)
where it is shown that the upper  $\om_{\kapa}^+$ 
 the lower  $\om_{\kapa}^-$ bands are parallel in $\kappa_{\perp}$ direction, shifted by the amount $\Om_{min}$. Once the  Fermi energy $\om_F$ is slightly
 larger than $\om_L$ ($\om_{F2}$ in Fig.~\ref{f5}(a)), a finite amount of states
determined by $\kappa_{\parallel}=0$ is excited across the pseudo-gap by the photon energy of $\Om = \Om_{min}$, producing a divergent optical response.
Overally, for  $\om_F$ well  within a pseudo-gap, the optical conductivity  behaves roughly as $\Om^{-3}$ over the entire interval between the points $\Om_{min}$ and $\Om_{max}$,
as it is shown in Fig.~\ref{f5}(b2) for several values of the $\om_L < \om_F < \om_U$ .\
\\

Case $\om_F > \om_U$: \\
In this case an additional feature appears in the optical conductivity  as can bee seen from Fig.~\ref{f5}(b3) plotted for several values of $\om_F$. Now, 
$-\Om_-$, which is positive Eq.~(\ref{op15}), is the limiting value, dividing the two different $\Om$-dependencies which arise  due to the square brackets in Eq.~(\ref{op16}). The rough dependence on $\Om$ can be summarized as :
\begin{eqnarray}\label{op19}
  &&  {\rm{Re}} \, \s^{(3)}(\Om) \sim \Om^{-1/2},  \hspace{3mm} \Om \gtrapprox \Om_{min} \nonumber \\
  &&  {\rm{Re}} \, \s^{(3)}(\Om) \sim \Om^{-2},  \hspace{3mm} \Om_{min}< \Om < -\Om_- \nonumber \\
 &&   {\rm{Re}} \, \s^{(3)}(\Om) \sim \Om^{-3},  \hspace{3mm} -\Om_-< \Om < \Om_{max}.
\end{eqnarray}
In Fig.~\ref{f5}(b3) a kink in ${\rm{Re}} \, \s^{(3)}(-\Om_-) $ is indicated by an arrow.
As $\om_F$ is further increased, the kink travels to the right.

\subsection{2D optical conductivity features}

Similar considerations apply for the 2D optical conductivity.\
\\

 Case $\om_F < \om_L$: \\
The 2D  optical conductivity which is shown in Fig.~\ref{f5}(c1) for several values of $\om_F$, has an onset of inter-band excitations at $\Om_-$. 
${\rm{Re}} \, \s^{(2)}(\Om) $ is  humped-like curve with a bit more pronounced amplitude then its 3D analog. 
The conductivity curve expands towards $\Om_{min}$, with increasing amplitude and spikes as $\om_F \to \om_L$.
Again, for $\om_F \ll \om_L$, which implies  $\Om_- \gg \Om_{min}$, the  position of maximum is roughly estimated to be at
 \begin{equation}\label{op20}
  \Om_0  \approx 5-\sqrt{1+24\om_F},
\end{equation}
denoted by the orange circles in Fig.~\ref{f5}(c1).
In the 2D case there is a square-root dependence ${\rm{Re}} \, \s^{(2)}(\Om) \propto \sqrt{\Om- \Om_-}/\sqrt{\Om_- - \Om_{min}}$  
for photon energies just above the onset of the inter-band excitation at $\Om_-$ as shown in Fig.~\ref{f5}(c1). 
The amplitude of the square-root increases as $\om_F \to \om_L$. 
Finally, for $\om_F = \om_L$,  Eq.~(\ref{op16}) reduces to the simple approximate expression 
 \begin{equation}\label{op21}
  {\rm{Re}} \, \s^{(2)}(\Om) \approx  \frac{\s^{(2)}}{\sqrt{2}} \frac{1}{\Om^2 \sqrt{\Om + \Om_{min}}},
\end{equation}
valid for energies $\Om$ not much larger than $\Om_{min}$.
For this particular value of Fermi energy, the spike of the optical conductivity is finite, located at $\Om_0 = \Om_{min}$, and has a height of 
 ${\rm{Re}} \, \s^{(2)}(\Om_0) = \s_0^{(2)}/2\Om_{min}^{5/2}$ (red dot in Fig.~\ref{f5}(c1)).

Although a singularity in the DOS is located at $\om_L$ (see Fig.~\ref{f2}), it does not imply the divergency of the optical conductivity as shown in Fig.~\ref{f5}(c1). 
The reason is the so-called ``bottle-neck`` effect that the upper band creates with its parabolic bottom and a constant DOS  (Figs. (\ref{f1}) and (\ref{f2}))
at energy $\om_U$. 
There is an infinite number of electron states
at $\om_L$  ready to be excited across the pseudo-gap to $\om_U$. However, at $\om_U$ there is only a finite number of states 
to accept those electrons. This is the origin of a finite  point-like dependence of optical conductivity Eq.~(\ref{op18}) at photon energy $\Om = \Om_{min}$.\
\\

Case $\om_L \leq \om_F \leq \om_U$:\\
As in the 3D case, here we also find a one over square-root divergency ${\rm{Re}} \, \s^{(2)}(\Om) \propto 1/\sqrt{\Om- \Om_{min}}$ for $\Om \gtrapprox \Om_{min}$
and $\om_F$ slightly above $\om_L$. However, unlike in the 3D case, there is no hump present. 
As $\om_F$ increases into the pseudo-gap, the square-root divergency is more and more pronounced. This also has its roots in the existence of a 
finite segment of the $\om_{\kapa}^-$ band that can be excited (vertically) to $\om^+_{\kapa}$ band 
 along the $\kappa_\perp$ direction at energy $\Om_{min}$. The higher the value of $\om_F$, the larger is the segment and so is the amplitude of the square-root 
divergency (Fig.~\ref{f5}(a)).
The optical conductivity is shown in Fig.~\ref{f5}(c2) for several values of $\om_L < \om_F < \om_U$.
Overally, for  $\om_F$ well  within the pseudo-gap, the optical response behaves roughly as $\Om^{-3}$ along the entire interval between the points $\Om_{min}$ and $\Om_{max}$.\
\\

Case $\om_F > \om_U$: \\
Here also, as it was in the 3D case,  $-\Om_-$ is the limiting value dividing the two different $\Om$-dependencies which arise  due to the expression within the square
brackets in Eq.~(\ref{op16}) (see Fig.~\ref{f5}(c3)). They are approximately given by 
\begin{eqnarray}\label{op22}
  &&  {\rm{Re}} \, \s^{(2)}(\Om) \sim \Om^{-1/2},  \hspace{3mm} \Om \gtrapprox \Om_{min} \nonumber \\
  &&  {\rm{Re}} \, \s^{(2)}(\Om) \sim \Om^{-2},  \hspace{3mm} \Om_{min}< \Om < -\Om_- \nonumber \\
 &&   {\rm{Re}} \, \s^{(2)}(\Om) \sim \Om^{-3},  \hspace{3mm} -\Om_-< \Om < \Om_{max}.
\end{eqnarray}
At the particular point,  ${\rm{Re}} \, \s^{(2)}(-\Om_-)$ has a small spike at energy $-\Om_-$, as pointed by an arrow  in Fig.~\ref{f5}(c3). 
As $\om_F$ is further increased, the kink (spike) travels to the right, while the overall conductivity curve moves to the left. This movement of the curve to the left is in contrast to the 3D case where the optical conductivity curve tends to move to the right as $\om_F$ is increased (see Figs. \ref{f5}(c3) and (b3)).

\subsection{Validity of the two-band model in the calculation of optical response}

The optical excitations in the two-band model are limited by the minimal $\Om_{min}$ and maximal $\Om_{max}$ value of the incoming photon.
$\Om_{max}$, which is equal to $\omega_T$, corresponds to the excitation as shown in Fig.~\ref{f5}(a). The question arises to which value of the Fermi energy, call it the critical Fermi energy $\om_F^c$, can we fill the second $\om_{\kapa}^+$ band so that electron excited from $\om_F^c$ hits the hypothetical third band (dashed line in Fig.~\ref{f5}(a)) when absorbing the
photon with the highest allowed energy $\Om_{max}$?
Once being above the energy $\om_F^c$, the two-band description of the optical conductivity is not sufficient and the third band has to be included into the consideration when 
calculating $\s^{(d)}(\Om)$. 

This scenario is relevant under the  assumption that higher Fourier components 
of the crystal potential are finite. This also implies that the  inter-band current matrix elements Eq.~(\ref{op5}) between the second and the third band are finite,
allowing the single-particle optical transitions.

The numerical estimate  for $\om_F^c$ can be done easily by taking the non-perturbative free electron dispersion of the third band $\e_{\kk-2\Q}$, shift it 
and express it in dimensionless units as described in Sec.~\ref{ham} to get $\om_{\kapa}$. Then we search  for the  wave vector $\kapa$
for which $\om_{\kapa} - \om^+_{\kapa} = \om_T$.
Neglecting $\eta$ and any other equivalent gap parameter in this procedure we obtain $\kappa \approx 1 -\om_T/8$. This in turn gives 
\begin{equation}
 \om_F^c \approx \left( 2-\om_T/8 \right)^2 \approx 2.25,
\end{equation}
where we have taken $\om_T \approx 4$.
Thus the result is that the optical response of the two-band model, as presented in Sec.~\ref{optika-kar}, is correct for the Fermi energy within the interval 
$\om_F \in (\om_B, \om^c_F)$. 
For $\om_F > \om_F^c$, a third band has to be taken into account for calculating the optical conductivity.

\section{Signatures of the topologically reconstructed Fermi surface}

In order to understand the signature of the Fermi surface reconstruction "seen" by the optical probe, let us briefly summarize the onset of this process in physical systems. The systems that spontaneously undergo the topological reconstruction of its Fermi surface do so by lowering the total energy with respect to the initial system without reconstruction. Usually, the spontaneous self-consistent periodic potential, that initiates the reconstruction, is some form of the density wave (charge or spin). Here, mainly two scenarios can take place: (1) the "nesting scenario" in which parts of the Fermi surface, with special open topology, get mapped to each other by a single wave vector, fully (or to the great extent) gaping the Fermi surface thus causing the corresponding density wave instability \cite{Gruner1988}; (2) the "touching scenario" in which the closed Fermi surfaces very slightly overlap each other leading to the formation of the pseudo-gap which lowers the band energy as predicted and well described 
in Refs. \cite{Kad2018, Spaic202}.
It is also worth mentioning the type of topological reconstruction of the Fermi surface triggered (due to some external influence such as pressure) by the inter-band instability of the initial system, consisting of distant parabolic valence and conduction band, in which the latter are related by a finite wave vector and reconstructed in the way to form the self-consistent "exciton band" which in turn lowers the overall energy of the system \cite{Kohn}.\

In this work  we analyze the optical response within the intra-band "touching scenario".  Any scenario of electron-driven instability is closely related to the Fermi wave vector, which, in turn, determines the wave vector of the density wave. It turns out that in such systems reconstruction of the Fermi surface takes place in such a way that the optimal Fermi energy "falls" in the pseudo-gap, i.e. $\omega_F \in \{ \om_L , \om_U \}$, between lower and upper Lifshitz point (closer to the upper one at $\omega_U$). Therefore, to change the Fermi energy significantly for the sake of optical probing does not seem to be feasible since it would compromise the stability of the system presented as such in this work. Nevertheless, in real systems it can be changed to some extent. The reason why is the crystal lattice potential (for simplicity not considered within the model under consideration). It permits the reconstruction with the wave vector which does not deviate from commensurability with the (reciprocal) crystal lattice to the great extent (see the charge ordering in the high-T$_c$ superconducting cuprates for example \cite{Keimer2015}). In that sense we may say that the density wave ordering with reasonably small adjustments of the Fermi energy, small compared to $\eta$, are protected by the crystal symmetry.\ 

Concerning the intra-band transport, which is a good probe once we can change the Fermi energy, the intra-band coefficients calculated in this paper are therefore evaluated only at one particular value of the Fermi energy, or its close vicinity. 
However, combining the knowledge about the Drude weight, which is finite, and Seebeck which is either zero (3D), or negative (2D), some information about the reconstructed phase can be obtained.\

Optical probe may give a better insight. As noted in the last section, we are interested in the shape of the optical conductivity when the Fermi energy lies inside the pseudo-gap, exactly given by Eq.~(\ref{op16}). Analysing the experimentally measured optical conductivity results, a hump-like feature, such as in  Figs.~\ref{f5} (b1) and (c1), or a kink-like one in Figs.~\ref{f5}(b3) and (c3), indicate that $\omega_F$ lies outside the pseudo-gap and hence we are not dealing with the topologically reconstructed Fermi surface.

Finally it is worth mentioning that, although there are some similarities in the shape of the inter-band conductivity (see Fig.\ref{f5})  with the inter-band conductivity of the 1D density wave condensate with finite and and infinite mass reported in \cite{Lee1974}, the process studied in this paper is of entirely different nature. In the afore-mentioned paper the density wave contribution to the conductivity is studied in the fully gapped 1D Fr\"{o}hlich system, while in this work we deal with the 2D/3D mostly metallic system in the pseudo-gap regime.  

\section{Conclusions}

In this paper we calculate the main zero-temperature intra-band charge transport coefficients and the real part of the (inter-band) optical conductivity for the 2D and 3D metallic system in which initially closed Fermi surface of the parabolic one-electron band undergoes the topological reconstruction into an open one. The topological reconstruction of the Fermi surface presumably takes place due to the instability of the initial electron band, with electron-phonon or electron-electron interaction, with respect to the spontaneous formation of the charge density wave with such a wave vector that relates the initial Fermi surfaces within so-called "touching scenario" resulting in formation of the saddle point in the lower and elliptical point in the upper newly formed electron band dispersion, and opening of a pseudo-gap in electron spectrum between them. In turn, the total energy of electron band is lowered, consequently stabilizing the density wave if the coupling constant of interaction is large enough  \cite{Kad2018, Kad2019, Spaic202}. Such a scenario may explain an onset of the charge density wave ground state accompanied with such a reconstruction of the Fermi surface, for example in the high-T$_c$ superconducting cuprates \cite{Keimer2015}, or certain intercalated graphite compounds \cite{Rahnejat2011}.\  

In order to track a signature of the specific reconstruction of the Fermi surface experimentally (for example the absence of the Hall effect in the 2D net of the closed Fermi surfaces reconstructed by a biaxial charge density wave as predicted in \cite{Kadigrobov2021}), we calculate the above-mentioned transport properties adopting, in this work, the two-band model. 
The electron density of states after the Fermi surface reconstruction, clearly exhibiting the pseudo-gap with logarithmic van Hove singularity in the 2D and Heaviside step discontinuity in the 3D system at the energy of the saddle point in the spectrum, has a profound impact on the calculated transport properties: the effective concentration of carriers taking part in electric conductivity, the Seebeck coefficient and the real part of the optical conductivity.\    

In terms of the Cartesian components with respect to the reconstruction wave vector, the perpendicular effective number of carriers shows no significant change, with respect to the total concentration of electrons, in both 2D and 3D case. The effective number of carriers addresses the problem of effective electron mass in the band reconstructions process, i.e. passing through the pseudo-gap as the Fermi energy is changed. On the other hand, the longitudinal component shows significant deviations form the total concentration in terms of reduction of concentration of electrons, more pronounced in 2D than in 3D, especially in the region of the pseudo-gap.
The signatures of the band reconstruction are even more striking in thermopower. The perpendicular component of the Seebek coefficient, similar to the effective electron concentration, does not show the significant deviation from the free electron value as the Fermi energy is changed. However, the parallel component attains negative value along the pseudo-gap, with large spikes at the energies of saddle point and elliptical point in electron band, in the 2D case, while in the 3D case the parallel Seebek coefficient is zero, indicating the absence of corresponding thermopower for Fermi energies within the pseudo-gap.\ 

The real part of the optical conductivity, within the framework of vanishing inter-band relaxation approximation, is presented in the closed form as a function of an incident photon energy and position of the Fermi energy for 2D and 3D system. We present all relevant features of the opical conductivity with respect to characteristic energies: minimal and maximal single-electron energy, and energies of saddle and elliptical points in the reconstructed electron band (so-called lower and upper Lifshitz points between which the pseudo-gap is spanned). For the Fermi energy within the interval below the lower Lifshitz point, optical conductivity is a finite hump-like curve in both 2D and 3D system. However, for the Fermi energy within the interval above the lower Lifshitz point, the optical conductivity attains a one over square root energy dependence for the incident photon energies above the pseudo-gap threshold (width of the pseudo-gap), also in both 2D and 3D system. Specificity of the 2D system is seen from comparison of the electron DOS and optical conductivity, where we see that the divergency in DOS does not imply the divergency in optical conductivity due to the so-called ``bottle-neck'' effect (due to the logarithmic van Hove singularity there is an infinite number of electrons to be excited into the finite number of available states). Futhermore, this implies that the oversimplified expression for real part of optical conductivity, involving the joint DOS \cite{Rukelj20211}, is not applicable here.

\section{Acknowledgments}
This work was supported by the Croatian Science Foundation, project IP-2016-06-2289, and by the QuantiXLie Centre of Excellence, a project cofinanced by the Croatian Government and European Union through the European Regional Development Fund - the Competitiveness and Cohesion Operational Programme (Grant KK.01.1.1.01.0004). The authors are thankful to dr. {\it I.K.} for the everlasting inspiration.\\

\appendix

\section{DOS mathematics}\label{apa}

We explicitly solve the $d = 3$ case.  The $d = 2$ follows accordingly. We investigate the $s=+$ case first. The Eq.~(\ref{dos4}) is 
\begin{eqnarray}\label{a1}
&& G_3(\om) \propto   \int_0^1   d\ka_{\parallel}  \, \Theta \left( \sqrt{\om  - \kappa_\parallel^2 - 1 - \sqrt{ 4 \kappa_\parallel^2 + \eta^2 }} \, \right) \nonumber \\
\end{eqnarray}
where we assumed the square root is real. Hence the sub-root function has to be positive. We solve this irrational inequality under the root 
to determine the boundaries of $\kappa_\parallel$ for which it holds
\begin{equation}\label{a2}
 \om >  \kappa_\parallel^2 + 1 + \sqrt{ 4 \kappa_\parallel^2 + \eta^2 } \Rightarrow  \kappa_\parallel \in \la 0, A_-(\om)  \ra
\end{equation}
where 
\begin{equation}\label{a3}
A_\pm(\om) = \sqrt{1+\om \pm \sqrt{4\om + \eta^2}} 
\end{equation}
Correspondingly, the $\Theta$-function in (\ref{a1}) can be written in a few alternative ways
\begin{eqnarray}\label{a4}
&&  \Theta \left( \sqrt{\kappa_\parallel \left(\kappa_\parallel -  A_-(\om)  \right)} \right) =  \Theta \left(\kappa_\parallel \left(  \kappa_\parallel -  A_-(\om)  \right) \right)  \nonumber \\
&& =  \Theta \left( \kappa_\parallel  \right) \Theta  \left(  A_-(\om)  - \kappa_\parallel   \right) =  \Theta  \left(  A_-(\om)  - \kappa_\parallel   \right) 
\end{eqnarray}
The last line in Eq.~(\ref{a4}) follows because $ \kappa_\parallel  > 0$.
 The necessary condition for  $A_-(\om)$ to be real gives $\om > 1+ \eta = \om_U$ while for it to be  $A(\om) < 1$ yealds $\om < 2+ \sqrt{4+\eta^2} = \om_T$.
The $\Theta$-function (\ref{a4}) once inserted back in (\ref{a1}) change the upper limit of integration. For those $\om$ that give $A(\om) < 1$, integral (\ref{a1}) becomes  
\begin{equation}\label{a5}
 G_3(\om) \propto    \int_0^{A_-(\om) }   d\ka_{\parallel}, 
\end{equation}
while for $\om > \om_T$ it stays 
\begin{equation}\label{a6}
G_3(\om) \propto   \int_0^1   d\ka_{\parallel} . 
\end{equation}
Similar reasoning applies for the $s=-$ case. 
Here the irrational inequality for the boundaries of $\kappa_\parallel$ gives
\begin{equation}\label{a7}
 \om >  \kappa_\parallel^2 + 1 - \sqrt{ 4 \kappa_\parallel^2 + \eta^2 } \Rightarrow  \kappa_\parallel \in \la A_-(\om), A_+(\om)  \ra.
\end{equation}
Since $A_+(\om) > 1,  \, \forall  \, \om $, we focus only on $A_-(\om)$ which is  $A_-(\om) < 1$ for  $\om > 2- \sqrt{4+\eta^2} = \om_B$
and $\om < 1-\eta = \om_L$.
Hence using the same recipe as in Eq.~(\ref{a4}) the $\Theta$-function can be written as 
\begin{eqnarray}\label{a8}
&&  \Theta \left( \sqrt{\om  - \kappa_\parallel^2 - 1 + \sqrt{ 4 \kappa_\parallel^2 + \eta^2 }} \, \right)  \nonumber \\
&& \equiv  \Theta \left( \kappa_\parallel -A_-(\om)  \right) \Theta  \left(  A_+(\om)  - \kappa_\parallel   \right) 
\end{eqnarray}
Now, the $\Theta$-function (\ref{a8}) changes in the integral (\ref{a1}) only the lower limit of integration. For those $\om$ that give $A_-(\om) < 1$, integral 
(\ref{a1}) becomes  
\begin{eqnarray}\label{a9}
&& G_3(\om) \propto   \int_{A_-(\om)}^1   d\ka_{\parallel}, 
\end{eqnarray}
while for $\om > \om_L$ it stays 
\begin{eqnarray}\label{a10}
&& G_3(\om) \propto  \int_0^1   d\ka_{\parallel}. 
\end{eqnarray}
Collecting the cases of different $\om$ values alongside with the  integration limits we come to  Eq.~(\ref{dos7}).

\bibliography{zadnja-dr}

\end{document}